\documentclass[aps,pre,twocolumn,superscriptaddress,nobibnotes]{revtex4-1}
\usepackage{graphicx}
\usepackage{amsmath}
\usepackage{dcolumn}
\usepackage{bm}
\usepackage{ulem}
\usepackage{times}
\usepackage{amssymb}
\usepackage{epsfig}

\newcommand{\lcal}{\mathcal{L}}

\begin{document}
\title{Relaxation Patterns in Supercooled Liquids from Generalized Mode-Coupling Theory}

\author{Liesbeth M.~C.~Janssen}
\email[Electronic mail: ]{lmj2130@columbia.edu}
\affiliation{Department of Chemistry, Columbia University, 3000 Broadway, New York, New York 10027, USA}
\author{Peter Mayer}
\altaffiliation[Current address: ]{ThinkEco, Inc., 494 8th Avenue, New York, NY 10001, USA}
\affiliation{Department of Chemistry, Columbia University, 3000 Broadway, New York, New York 10027, USA}
\author{David R.~Reichman}
\email[Electronic mail: ]{drr2103@columbia.edu}
\affiliation{Department of Chemistry, Columbia University, 3000 Broadway, New York, New York 10027, USA}

\date{\today}

\begin{abstract}

The mode-coupling theory of the glass transition treats the dynamics of
supercooled liquids in terms of two-point density correlation functions.
Here we consider a generalized, hierarchical formulation of schematic mode-coupling
equations in which the full basis of multipoint density correlations is taken
into account. By varying the parameters that control the effective
contributions of higher-order correlations,  we show that infinite hierarchies
can give rise to both sharp and avoided glass transitions. Moreover, small
changes in the form of the coefficients result in different scaling behaviors
of the structural relaxation time, providing a means to tune the fragility in
glass-forming materials. This demonstrates that the infinite-order construct of
generalized mode-coupling theory constitutes a powerful and unifying framework
for kinetic theories of the glass transition.
\end{abstract}

\maketitle

The glass transition of liquids and dense colloidal suspensions represents one
of the most puzzling phenomena in all of condensed matter science.  Among the
various theories of glass formation proposed in the last few decades,
mode-coupling theory (MCT) has acquired a unique place in this area of
research. In particular, MCT can reproduce important features of the
time-dependent dynamics and relaxation of supercooled liquids using only static
information as input, making it essentially the only theory of glassy dynamics
based entirely on first principles \cite{gotze:09, berthier:11}.

The main quantity of interest in MCT is the two-point density correlation
function $F(k,t) = N^{-1} \langle \rho_{\mathbf{-k}}(0) \rho_{\mathbf{k}}(t)
\rangle$, where $\mathbf{k}$ is a wavevector with magnitude $k$,
$\rho_{\mathbf{k}}(t)$ is the $\mathbf{k}$-th Fourier component of the spatial
density fluctuations at time $t$, $N$ is the number of particles, and the
brackets denote a canonical ensemble average. The equation of motion for
$F(k,t)$ contains a memory function that, within the standard MCT framework,
is assumed to be dominated by pair-densities. This allows one to factorize the
memory kernel, which is essentially a four-point density correlator, into a
product of two two-point correlators.  Applying additional Gaussian and
convolution approximations for the static multi-point correlation functions
subsequently yields a closed, self-consistent expression for the time evolution
of $F(k,t)$.

One of the successes of standard MCT is its ability to capture the ``cage
effect" responsible for an intermediate-time plateau in correlation functions
and the dramatic slow-down of dynamics upon decreasing the temperature or
increasing the density. The scaling properties of $F(k,t)$ associated with this
$\beta$-relaxation process are also accurately reproduced.  The power-law
divergence of the $\alpha$-relaxation time ($\tau$) predicted by MCT is
consistent with experiments and computer simulations, but only in the mildly
supercooled regime. A major drawback of the theory is that it predicts an ideal
glass transition at relatively high temperatures or low densities.
Furthermore, standard MCT cannot account for exponential (Arrhenius) or
super-Arrhenius dependences of $\tau$ in the deeply supercooled regime, and
thus cannot describe distinct ``strong" and ``fragile" glass-forming behaviors,
respectively \cite{angell:95}. 

In an effort to account for activated processes that round off the ideal MCT
transition, Das and Mazenko \cite{das:86} and G\"{o}tze and Sj\"{o}gren
\cite{gotze:87} extended standard MCT by considering additional perturbative
couplings to current modes. Although this approach can improve the predicted
behavior of standard MCT in the case of strongly supercooled liquids, it does
not apply to (hard-sphere) systems undergoing Brownian motion, 
where current modes play no substantial role. Indeed,
several recent theoretical and computational studies have disputed the
importance of density-current mode coupling close to the glass transition on
general grounds \cite{cates:06, andreanov:06, szamel:13}.

An alternative improvement to standard MCT, referred to as generalized MCT
(GMCT), was first introduced by Szamel in 2003 \cite{szamel:03}.  GMCT relies
on the fact that the exact time evolution of four-point density correlations is
governed by six-point correlation functions, which in turn are controlled by
eight-point correlations, and so on. This makes it possible to delay the
factorization approximation for the memory kernel to a later stage.  Numerical
studies employing second- and third-order truncations have shown that GMCT
indeed systematically improves the predicted MCT transition temperature (or
volume fraction), implying that higher-order correlations account for at least
some features ignored by standard MCT in the deeply supercooled regime
\cite{szamel:03,wu:05}.

More recently, two of us extended the GMCT approach to \textit{infinite} order
using a simplified schematic model based on the form of the microscopic
equations of motion, allowing the factorization closure to be rigorously
avoided \cite{mayer:06}. The general form of this infinite hierarchy reads 
(see Appendix \ref{app:kGMCT})
\begin{equation}
\label{eq:sGMCThierarchy}
\dot{\phi}_n(t) + \mu_n \phi_n(t) + \lambda_n \int_0^t \phi_{n+1}(\tau)
\dot{\phi}_n(t-\tau) d\tau = 0,
\end{equation}
where the functions $\phi_n(t)$ represent normalized $2n$-point density
correlators ($n \in \mathbb{N}$), $\mu_n$ are generalized bare frequencies, and
the $\lambda_n=\lambda_n(\Lambda)$ parameters play the role of generalized inverse-temperature-like
coupling constants. Here $\Lambda$ can be thought of as the control parameter of the transition,
e.g.\ inverse temperature or volume fraction.
In arriving at the form of Eq.\ (\ref{eq:sGMCThierarchy}),
we have employed Gaussian and convolution approximations for the static
correlations, included only diagonal contributions to the memory functions, and
treated all wavevectors on an equal footing.
These approximations are all similar to those
employed in standard MCT. We emphasize that Eq.\ (\ref{eq:sGMCThierarchy}) is based on a 
fully microscopic theory, as detailed in Appendix \ref{app:kGMCT}.
In Ref.\ \cite{mayer:06} we
considered a simple hierarchy with $\mu_n=n$ and $\lambda_n=\Lambda$, a constant, and found
that it admits an analytic solution which is characterized by a continuously
growing, exponentially diverging relaxation time. This result is to be
contrasted with finite-order GMCT, which always predicts a power-law divergence
at a sharp MCT transition.  The inclusion of all multipoint dynamical
correlations thus provides a means to strictly remove the sharp MCT transition
and convert power-law divergences of $\tau$ into exponentially varying forms.

In this article, we further elaborate on the infinite schematic GMCT framework
and show that, by considering more general forms for the $\mu_n$ and 
$\lambda_n(\Lambda)$ parameters of the hierarchy, 
GMCT can account for a vast wealth of relaxation patterns in supercooled liquids.
More explicitly, we demonstrate that infinite GMCT hierarchies can reproduce the many
features of the standard-MCT-based $F_2$ model--which is characterized by a sharp MCT
transition and power-law relaxation \cite{leutheusser:84,bengtzelius:84}--but can also reveal novel relaxation
patterns beyond those predicted by standard MCT.
We also discuss how both strong and fragile relaxation motifs can emerge
within infinite GMCT hierarchies devoid of sharp transitions by tuning the
$n$-dependence of the $\lambda_n$-parameters.  This constitutes the first
kinetic-theory-motivated framework that can account for different fragilities in glass-forming
materials.

Let us first consider some general features of Eq.\ (\ref{eq:sGMCThierarchy})
and its solutions $\{ \phi_n(t) \}$. An important quantity in our present
discussion is the $\alpha$-relaxation time for the $n$-th level density
correlator, which we define as
\begin{equation}
\label{eq:taudef}
\tau_n = \int_0^{\infty} \phi_n(t) dt = \hat{\phi}_n(s=0).
\end{equation}
Here $\hat{\phi}_n(s)$ is the Laplace transform of $\phi_n(t)$, defined by
$\hat{\phi}_n(s) = \lcal\{\phi_n(t) \} = \int_0^{\infty} \phi_n(t) e^{-st} dt$.
By iterating the Laplace-transformed solution of Eq.\ (\ref{eq:sGMCThierarchy}) for $s=0$, we obtain
(see Appendix \ref{app:kGMCT})
\begin{equation}
\label{eq:phin_s=0it}
\tau_n \simeq \frac{1}{\mu_n} 
          \sum_{m=0}^{\infty} \prod_{i=0}^{m-1} \frac{\lambda_{n+i}}{\mu_{n+1+i}}. 
\end{equation}
Instead of considering the $\alpha$-relaxation time, one may also characterize
the glass transition in terms of the long-time limit of $\phi_n(t)$. 
We define this long-time limit as $q_n = \lim_{t \to \infty} \phi_n(t) = \lim_{s \to 0} s
\hat{\phi}_n(s)$, which can be written explicitly as (see Appendix \ref{app:kGMCT})
\begin{equation}
\label{eq:qn_it}
\frac{1}{q_n} = 1 + \frac{\mu_n}{\lambda_n} \frac{1}{q_{n+1}} 
\simeq 
\sum_{m=0}^{\infty} \prod_{i=0}^{m-1} \frac{\mu_{n+i}}{\lambda_{n+i}}.
\end{equation}
Equations (\ref{eq:phin_s=0it}) and (\ref{eq:qn_it}) are our general expressions for the relaxation time and
long-time limit of $\phi_n(t)$, respectively, as governed by the infinite hierarchy of Eq.\ (\ref{eq:sGMCThierarchy}).

The convergence behavior of the general expressions (\ref{eq:phin_s=0it}) and 
(\ref{eq:qn_it}) can already reveal important information on the type of transition
contained in the hierarchy. For an MCT-like transition, there exists a critical point 
$\Lambda=\Lambda_c$ above which the $\phi_n(t)$ no longer decay to zero. This nonzero long-time 
limit $q_n$ may grow continuously (type-A transition) or discontinuously (type-B transition) 
as a function of the control parameter $\Lambda$. If the transition is completely avoided, the
relaxation time grows continuously but ultimately leads to full relaxation of
the correlation functions ($q_n=0$) for all finite $\Lambda$. 
One may verify that 
the series for $\tau_n$ converges 
if $\lim_{n \to \infty} \lambda_n/\mu_{n+1} < 1$, and $1/q_n$ converges 
if $\lim_{n \to \infty} \mu_n/\lambda_n < 1$. A type-A transition is characterized
by a diverging series for both $\tau_n$ and $1/q_n$ at $\Lambda=\Lambda_c$,
while a type-B transition has a diverging $\tau_n$ series and converging $1/q_n$ series. 
Conversely, for a rigorously avoided transition, $\tau_n$ converges and $1/q_n$ diverges
for all $\Lambda$.
Thus, depending on the asymptotic behavior of the $\{\mu_n, \lambda_n\}$ coefficients,
the GMCT framework can account for all of these physically distinct phenomena.
This is one of the key results of this work: by making a suitable choice for 
$\mu_n$ and $\lambda_n$, we can generate arbitrary types of transitions and,
by virtue of Eqs.\ (\ref{eq:phin_s=0it}) and (\ref{eq:qn_it}), arbitrary scaling 
behaviors of the relaxation time and long-time limit. The chosen set of $\{\mu_n, \lambda_n\}$ 
coefficients subsequently determines the full hierarchy and all its time-dependent 
solutions $\{\phi_n(t)\}$.

In the remainder of this paper, we shall focus on some explicit examples of
the general hierarchy (\ref{eq:sGMCThierarchy}) and 
restrict our discussion to the two-point density correlator $\phi_1(t)$, i.e.\
$n=1$. All other correlation functions ($n>1$) appear only as
generalized memory functions for $\phi_1(t)$.
We first consider a class of hierarchies that exhibit MCT-like, type-B transitions
but are fundamentally distinct from the standard-MCT $F_2$ model.
Note that the $F_2$ model is essentially the lowest-order truncation of
GMCT with closure $\phi_2(t) = \phi_1^2(t)$ and $\lambda_1 = 4\mu_1 \Lambda$
\cite{leutheusser:84}.  We start with a relatively simple infinite hierarchy of
the form $\mu_n = n$ and $\lambda_n = \Lambda (n+c)$, where $c \geq 0$.  The
choice $\mu_n = n$ follows naturally from the
microscopic derivation of Eq.\ (\ref{eq:sGMCThierarchy}), provided that no
explicit distinction is made between different wavevectors \cite{mayer:06}.  
The functional form $\lambda_n = \Lambda (n+c)$ implies that the coupling parameters
$\lambda_n$ also grow linearly with $n$, and will remain on the same order of magnitude
as the frequencies $\mu_n$ for all levels $n$.
We obtain
for the relaxation time $\tau_1$ [see Eq.\ (\ref{eq:phin_s=0it})]
\begin{equation}
\tau_1 = \sum_{m=0}^{\infty} \left( \prod_{i=1}^{m} \frac{i+c}{i+1} \right) \Lambda^m 
       = \frac{1}{c\Lambda} \left[ \frac{1}{(1-\Lambda)^c} -1 \right],
\end{equation}
which diverges, to leading order, as $\tau_1 \sim (1-\Lambda)^{-c}$ near the
critical point $\Lambda_c=1$. Thus, 
our infinite hierarchy with parameters $\{ \mu_n=n, \lambda_n = \Lambda (n+c) \}$
predicts a \textit{power-law} divergence of the $\alpha$-relaxation time,
similar to the type-B transition in standard MCT. In fact, setting $c$ equal to
the standard-MCT $\alpha$-exponent $\gamma\approx 1.765$ \cite{leutheusser:84} yields exactly the
same power-law behavior, implying that certain features of the schematic $F_2$
MCT model can be accurately reproduced by a very simple infinite GMCT
hierarchy. 

Let us now consider the long-time limit of the density-density correlation
function $\phi_1(t)$ within this class of hierarchies. From Eq.\ (\ref{eq:qn_it}), 
we find
\begin{equation} 
\frac{1}{q_1} = \sum_{m=0}^{\infty} \left( \prod_{i=1}^{m} \frac{i}{i+c} \right)
                                    \left( \frac{1}{\Lambda} \right)^m.
\end{equation}
Explicitly, for a hierarchy with $c=0$, the inverse plateau height scales with
$\Lambda$ as $1/q_n = \Lambda/(\Lambda-1)$ for all $n$, the case $c=1$ gives
rise to logarithmic scaling, $1/q_1 = \Lambda \log[\Lambda/(\Lambda-1)]$, and
for $c=\gamma$ we have, for $\Lambda \gtrsim 1$,  
$1/q_1 = \frac{\gamma}{\gamma-1} \left( 1+\frac{\pi(\gamma-1)}{\sin(\pi\gamma)} (\Lambda-1)^{\gamma-1}
+ \mathcal{O}(\Lambda-1) \right)$
(see Appendix \ref{app:sharp}). 
In all cases, $q_1 > 0$ for $\Lambda > 1$, confirming that these GMCT hierarchies 
induce something akin to type-B
transitions at the critical point $\Lambda_c=1$ \cite{gotze:09}. The expressions for $q_1$
should, however, be contrasted with the $F_2$ model \cite{leutheusser:84}, which predicts a
\textit{square-root} scaling of the form $q^{F_2}_1 = (1 + \sqrt{1 -
1/\Lambda})/2$. Thus, standard-MCT-like hierarchies with
$\{\mu_n=n, \lambda_n = \Lambda (n+c)\}$ reveal entirely novel scaling 
behavior of the plateau height near the transition.

One may also ask whether it is possible to find an infinite hierarchy that
exhibits \textit{precisely} the same $\Lambda$-dependence as the $F_2$ model,
both with respect to $q_1$ and $\tau_1$. A comparison between the Taylor 
series of 1/$q^{F_2}_1$ [Eq.\ (\ref{eq:TaylorqF2})] and Eq.\ (\ref{eq:qn_it}) 
reveals that such 
an exact mapping requires 
\begin{equation}
\label{eq:F2match_lambdan_gamman}
\frac{\lambda_n}{\mu_n} = \frac{2n+2}{2n-1} \Lambda.
\end{equation}
By now fitting the $\mu_n$ parameters such that they also reproduce the power-law relaxation
time of the $F_2$ model (see Appendix \ref{app:sharp}), we obtain
\begin{equation}
\label{eq:F2match_gamman}
\mu_n = \prod_{i=1}^{n-1} \frac{(2i+2)(i+b)}{(2i-1)(i+a)},
\end{equation}
with $a \approx 0.52726$ and $b \approx -0.23772$ determined numerically from the fit.
Thus, an infinite GMCT hierarchy with coefficients
(\ref{eq:F2match_lambdan_gamman}) and (\ref{eq:F2match_gamman}) represents a 
numerically motivated approximation to the $F_2$ model.

\begin{figure}[!htb]	
  \begin{center}
    \includegraphics[width=8.5cm]{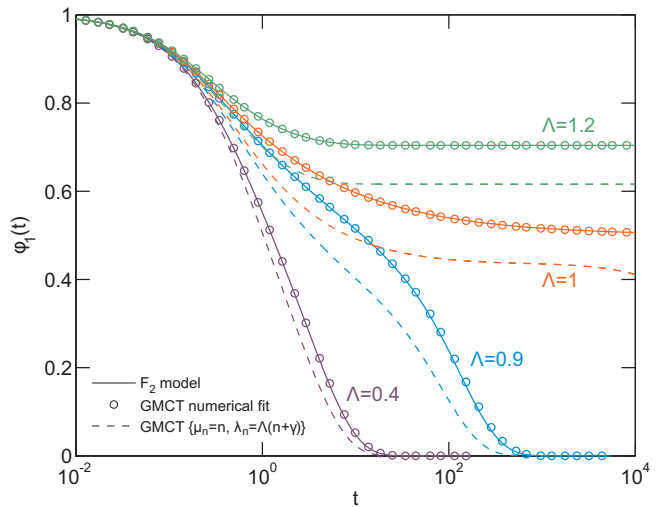}
  \end{center}
  \caption{
  \label{fig:F2match} 
  Density-density correlation functions $\phi_1(t)$ for the $F_2$ model (solid
  lines), the fitted GMCT hierarchy with coefficients
  (\ref{eq:F2match_lambdan_gamman}) and (\ref{eq:F2match_gamman}) (circles), and
  the GMCT hierarchy with $\{\mu_n =n, \lambda_n=\Lambda(n+\gamma)\}$ (dashed lines),
  calculated for $\Lambda=0.4, 0.9, 1$ and 1.2. 
   }
\end{figure}

Figure \ref{fig:F2match} compares the functions $\phi_1(t)$ obtained from the
various MCT-like hierarchies discussed above, and from the 
exact $F_2$ model \cite{leutheusser:84}, for different values of $\Lambda$.
The GMCT hierarchies were truncated after
$N=10000$ levels by exponential closure [$\phi_N(t) = \exp(-Nt)$], which is
amply sufficient to ensure convergence of the numerical solutions.
All data were obtained using the time-integration algorithm of Fuchs \textit{et al.}\ \cite{fuchs:91}.
For all values of $\Lambda$ considered, the hierarchy with $\{\mu_n=n, \lambda_n = \Lambda (n+\gamma)\}$ 
exhibits clear deviations from the $F_2$ model, yet exhibits exactly the same $\tau_1$ power-law scaling.
The numerically fitted GMCT hierarchy [Eqs.\ (\ref{eq:F2match_lambdan_gamman}) and (\ref{eq:F2match_gamman})],
on the other hand, 
show almost perfect agreement with the exact $F_2$ result, reproducing the complete time
dependence of $\phi_1(t)$ over all 6 decades of time. 
This is a remarkable and
highly non-trivial result: by fitting only the \{$\lambda_n,\mu_n$\} parameters to a
certain plateau height and relaxation time, we capture all
qualitative and quantitative features of a finite-order schematic MCT model,
both as a function of time and of $\Lambda$. We emphasize that no information of the short dynamics
was included in the fit, and yet this time domain is also accurately described.
We thus argue that GMCT is indeed more general in the sense that it can successfully reproduce 
the predictions of standard MCT, but can also predict entirely novel transitions and relaxation patterns
beyond those contained in the standard MCT scenario.

We now turn our attention to a different family of infinite GMCT hierarchies that are devoid of sharp
MCT-like transitions. 
We will focus on
hierarchical equations of the form $\mu_n = n$ and $\lambda_n = \Lambda
n^{1-\nu}$, with $\Lambda, \nu > 0$.  Note that the infinite hierarchy with
$\mu_n = n$ and $\lambda_n = \Lambda$, which has already been discussed
in Ref.\ \cite{mayer:06}, is a special case of this type of hierarchy. For
this choice of parameters, 
the relaxation time
becomes, after some manipulation (see Appendix \ref{app:avoided})  
\begin{equation}
\label{eq:tau_asymp}
\tau_1(\Lambda) 
\sim \frac{(2\pi)^{(1-\nu)/2}}{\nu^{1/2}} \Lambda^{-\frac{\nu+1}{2\nu}} \exp({\nu \Lambda^{1/\nu}}),
\end{equation}
where we have assumed that $\Lambda \gg 1$. This assumption holds in the deeply
supercooled regime, i.e.\ at asymptotically low temperatures. 
It is important to remark that, in contrast to the standard MCT prediction, the
relaxation time of Eq.\ (\ref{eq:tau_asymp}) does not diverge at any finite
$\Lambda$. This implies that there is no sharp MCT transition at any finite
$\Lambda$ for an infinite GMCT hierarchy of the form $\lambda_n = \Lambda
n^{1-\nu}$. Instead the relaxation time grows continuously with $\Lambda$, as
was already found in Ref.\ \cite{mayer:06} for the special case $\nu=1$.
It may be verified that the long-time limit of $\phi_1(t)$ for this class of
hierarchies also vanishes for all $\Lambda,\nu > 0$, confirming that the
transition is rigorously avoided.

\begin{figure}[!t]
  \begin{center}
    \includegraphics[width=8.5cm]{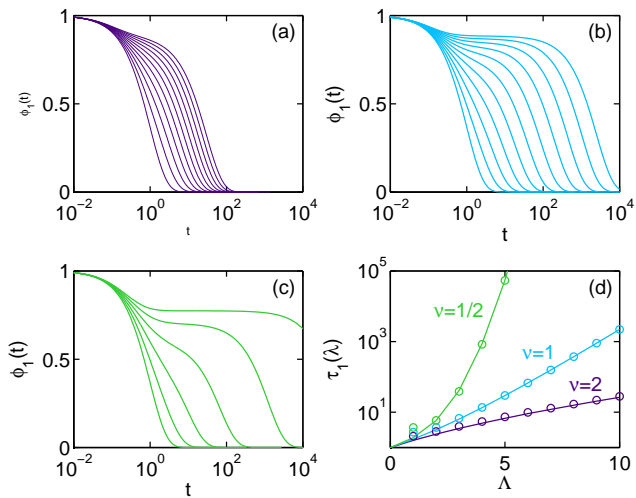} 
  \end{center}
  \caption{
  \label{fig:phit} 
  Solutions $\phi_1(t)$ of infinite hierarchies with $\mu_n=n$ and
$\lambda_n = \Lambda n^{1-\nu}$ for (a) $\nu = 2$ and $\Lambda = 0,1,\ldots
10$, (b) $\nu=1$ and $\Lambda = 0,1,\ldots 10$, (c) $\nu=1/2$ and $\Lambda =
0,1, \ldots 5$. The fastest decaying functions correspond to $\Lambda=0$.
Panel (d) shows the associated relaxation times $\tau_1(\Lambda)$: the solid
lines were obtained by numerical integration of $\phi_1(t)$ over time [Eq.\
(\ref{eq:taudef})], and the circles represent the analytical result of Eq.\ 
(\ref{eq:tau_asymp}).}
\end{figure}

Let us look at some explicit examples of these avoided GMCT transitions. 
For $\nu=1$ ($\lambda_n = \Lambda$), we recover the hierarchy of Ref.\
\cite{mayer:06}, with $\tau_1(\Lambda) \sim \exp(\Lambda)/\Lambda$.  The case
$\nu=2$ ($\lambda_n = \Lambda/n$) yields 
(see Appendix \ref{app:avoided}) 
$\tau_1 = I_1(2\sqrt{\Lambda})/\sqrt{\Lambda}$, where
$I_l$ is the modified Bessel function of the first kind.  For large $\Lambda$,
the relaxation time then behaves as $\tau_1 \sim (4\pi)^{-1/2} \Lambda^{3/4}
\exp(2\sqrt{\Lambda})$, in accordance with Eq.\ (\ref{eq:tau_asymp}). The
density correlation functions $\phi_1(t)$ for these two hierarchies, as well as
those for $\nu=1/2$, are shown in Fig.\ \ref{fig:phit} for various values of
$\Lambda$.  The data have been obtained from numerical integration of the
hierarchical equations using the algorithm of Ref.\ \cite{fuchs:91} with exponential
closure at $N=1000$.  One can see that, for fixed $\Lambda$, the correlation
functions decay more rapidly as $\nu$ increases, as predicted by Eq.\
(\ref{eq:tau_asymp}). 
This is the analog of the system become more
fragile with increasing $\nu$. 
The difference in fragility between hierarchies with
different $\nu$ is best observed by examining the relaxation times $\tau_1$ as
a function of $\Lambda$ [Fig.\ \ref{fig:phit}(d)].  These data were generated
by numerically integrating the $\phi_1(t)$ over time [Eq.\ (\ref{eq:taudef})].
It is clear that hierarchies with small $\nu$ give the most fragile behavior,
i.e.\ the relaxation time increases more dramatically with varying $\Lambda$ as $\nu$ approaches zero.
For comparison, we also show the analytical expression for $\tau_1$ [Eq.\
(\ref{eq:tau_asymp})] in Fig.\ \ref{fig:phit}(d); the agreement with the
numerical data is seen to be very good for $\Lambda > 1$.  As a final point,
we note that other features of $\phi_1(t)$ are also affected by $\nu$, e.g.\
the plateau height of $\phi_1(t)$ in the $\beta$-relaxation regime. In fact
it has been noted that strong glass formers generally have larger plateau values
compared to fragile ones \cite{mattsson:09}. 
The precise
characterization of these features will 
be discussed in future work.

In the deeply supercooled regime, the true relaxation time diverges as an
Arrhenius (exponential) or super-Arrhenius law, depending on the fragility of the system \cite{angell:95}. It
is well established that standard MCT cannot account for such fragilities, and
instead always predicts a power-law divergence of $\tau_1$. We find, however,
that GMCT hierarchies of the form 
$\{\mu_n = n$, $\lambda_n = \Lambda n^{1-\nu}\}$ ($\Lambda, \nu > 0$)
can account for different degrees of fragility depending on the value of $\nu$.
This is an important result: 
the $n$-dependence of the coupling strengths
$\lambda_n$ in infinite-order GMCT provides a means to tune the fragility of a
glass-forming system. 
While this finding is based on only a schematic
description of the dynamics, one may expect it to be preserved in a fully
microscopic version of GMCT, similar to how the qualitative features of the
schematic $F_2$ model are reproduced in $\textbf{k}$-dependent standard MCT.

Finally, we briefly elaborate on the physical interpretation of the various
GMCT hierarchies discussed in this work. First recall that the ratio 
$\lim_{n\rightarrow\infty} \lambda_n/\mu_{n+1}$ determines whether the transition
is sharp or avoided. For an avoided transition, 
the contributions of the higher-order memory kernels ultimately vanish at sufficiently
large $n$, i.e.\ the couplings $\lambda_n$ will become negligible compared to the bare 
frequencies $\mu_n$. Conversely, for sharp transitions, the $\lambda_n$ always remain 
on the same order of magnitude as the frequencies $\mu_n$ for all $n$. Thus, sharp 
MCT-like transitions contain significant contributions from \textit{all} terms 
up to $n\rightarrow\infty$; damping of the large-$n$ couplings will more strongly round off the MCT transition. 
For the class of avoided transitions studied here, with $\lambda_n=\Lambda n^{1-\nu}$, we
see that the $\lambda_n$ parameters decay more slowly with $n$ as $\nu$ decreases. This corresponds to
a higher degree of fragility, and hence fragile systems are governed by relatively large contributions
from higher-order dynamic correlations. 
In the limiting case $\nu=0$ we recover the linear hierarchy with $c=0$, and the avoided transition turns into
a sharp transition with (fragile) power-law relaxation. This is reminiscent of the MCT transition in the $F_2$ model,
which is generally believed to reflect a mean-field-like scenario. Thus, the `fragility'
parameter $\nu$ might possibly be regarded as a measure for the mean-field (or MCT-like)
character of the transition: the smaller the value of $\nu$, the more
MCT-like the nature of system. 
An alternative interpretation of $\nu$ follows from the notion that
fragility is linked to microscopic structure, with more fragile systems exhibiting larger variations
in short-ranged order than strong glass formers (see e.g.\ Ref.\ \cite{louzguine:11}). 
A stronger increasing dependence in $\lambda_n$
(i.e.\ smaller $\nu$) provides a means to increase the coupling between these structural motifs and
the dynamics. For strong systems, it is expected that such intricate structural effects play a much smaller
role in the relaxation dynamics. This could be captured in the GMCT hierarchy by rendering the coupling
terms $\lambda_n$ less important as the level $n$ increases, i.e. by increasing $\nu$.

In summary, we have presented a schematic generalized mode-coupling theory in
which dynamic multipoint density correlations are included through an infinite
hierarchy of coupled equations. 
Such a hierarchical framework can 
accurately capture the many features of the standard-MCT $F_2$ model, but can also give rise
to generalized new forms of glass transitions,
implying that there may be a purely dynamic origin for distinct glassy relaxation patterns.
Moreover, a suitable choice of the
coupling strengths of the higher-order correlations can lead to Arrhenius and
super-Arrhenius behavior of the $\alpha$-relaxation time, providing a means to
tune the degree of fragility with a single parameter. This represents the
first MCT-based theory that can account for different fragilities in
glass-forming materials.

\acknowledgments
LMCJ gratefully acknowledges support from the Netherlands Organization for
Scientific Research (NWO) through a Rubicon fellowship, and from IMI-NFG (NSF
grant DMR-0844014).  DRR ackowledges grant NSF-CHE 1213247 for support.

\appendix

\section{Microscopic and schematic GMCT equations}
\label{app:kGMCT}
In this Appendix, we provide information on the microscopic ($\mathbf{k}$-dependent) foundation
of our GMCT framework, and on the mathematical details of some of the
schematic ($\mathbf{k}$-independent) GMCT results discussed in the main text.

Our schematic GMCT equations are based on a fully microscopic theory 
that accounts for the dynamics of the diagonal $\mathbf{k}$-dependent $2n$-point density
correlation functions $\phi_n(t)$,
\begin{gather}
\phi_1^{(k)}(t)            =  
    \frac{\langle \rho_{\mathbf{-k}} \rho_{\mathbf{k}}(t) \rangle}{\langle \rho_{\mathbf{-k}} \rho_{\mathbf{k}} \rangle}, \nonumber \\  
\phi_2^{(k_1,k_2)}(t)      =  
    \frac{\langle \rho_{\mathbf{-k_1}} \rho_{\mathbf{-k_2}} \rho_{\mathbf{k_1}}(t) \rho_{\mathbf{k_2}}(t) \rangle}
    {\langle \rho_{\mathbf{-k_1}} \rho_{\mathbf{-k_2}} \rho_{\mathbf{k_1}} \rho_{\mathbf{k_2}} \rangle}, \nonumber \\
\phi_3^{(k_1,k_2,k_3)}(t)  =  
    \frac{\langle \rho_{\mathbf{-k_1}} \rho_{\mathbf{-k_2}} \rho_{\mathbf{-k_3}} \rho_{\mathbf{k_1}}(t) \rho_{\mathbf{k_2}}(t) \rho_{\mathbf{k_3}}(t) \rangle}
    {\langle \rho_{\mathbf{-k_1}} \rho_{\mathbf{-k_2}} \rho_{\mathbf{-k_3}} \rho_{\mathbf{k_1}} \rho_{\mathbf{k_2}} \rho_{\mathbf{k_3}} \rangle}, \nonumber \\
    \hdots 
\label{eq:phinorm}
\end{gather}
Each of these correlation functions is governed by an integro-differential equation with a memory kernel
containing $2(n+1)$-point correlators. Assuming Gaussian factorization for the \textit{static} correlations, 
we obtain for these memory kernels at level $n$,
\begin{eqnarray}
\label{eq:Knschematic}
K_1(k,t) &=& \frac{\rho k_{\rm{B}}T}{16m\pi^3} \int d\mathbf{q} |V_{\mathbf{q,k-q}}|^2
              S(q) S(|\mathbf{k-q}|) \times \nonumber \\
             & & \phi_2^{(q,k-q)}(t),
\nonumber \\
K_2(k_1,k_2,t) &=& \frac{\rho k_{\rm{B}}T}{16m\pi^3} \sum_{i=1}^2 \frac{\mu_{k_i}}{\mu_{k_1}+\mu_{k_2}}
                  \times \nonumber \\
              & & \int d\mathbf{q} |V_{\mathbf{q,k_i-q}}|^2 
                  S(q) S(|\mathbf{k_i-q}|) \times \nonumber \\
              & & \phi_3^{(q,k_1-q\delta_{i1},k_2-q\delta_{i2})}(t),
\nonumber \\
K_3(k_1,k_2,k_3,t) &=& \frac{\rho k_{\rm{B}}T}{16m\pi^3} \sum_{i=1}^3 \frac{\mu_{k_i}}{\mu_{k_1}+\mu_{k_2}+\mu_{k_3}}
                  \times \nonumber \\
               & &  \int d\mathbf{q} |V_{\mathbf{q,k_i-q}}|^2 
                  S(q) S(|\mathbf{k_i-q}|) \times \nonumber \\
               & &  \phi_4^{(q,k_1-q\delta_{i1},k_2-q\delta_{i2},k_3-q\delta_{i3})}(t), \nonumber \\
\hdots & & 
\end{eqnarray}
where for simplicity, and following Refs.\ \cite{szamel:03,wu:05}, we have retained only the diagonal dynamic contributions to the integral.
In Eq.\ (\ref{eq:Knschematic}), $\rho$ denotes the total density, $k_{\rm{B}}$ is the Boltzmann constant, $T$ is the temperature, $m$ is the
particle mass, $V_{\mathbf{q,k-q}}$ are static vertices, $S(q)$ is the static structure factor,
$\mu_{k_i} = \frac{k_{\rm{B}} T k_i^2}{m S(k_i)}$ is the bare frequency, and $\delta_{ij}$ represents the Kronecker delta function.
A detailed derivation of Eq.\ (\ref{eq:Knschematic}), as well as its application to a realistic microscopic system, 
will be provided in a forthcoming publication.
Following Ref.\ \cite{mayer:06}, we now drop the wavevector indices and treat
all wavevectors on an equal footing.  That is, $\phi_1^{(k)}(t) \mapsto
\phi_1(t)$, $\phi_2^{(k_1,k_2)}(t) \mapsto \phi_2(t)$, \dots, and
$\mu_{k_1}+\hdots+\mu_{k_n} \mapsto \mu_n$. Note that $\mu_n \propto n$ follows naturally
if no distinction is made between different $k$-values. We also replace $\frac{\rho
k_{\rm{B}}T}{16m\pi^3} \sum_{i=1}^n \int d\mathbf{q}
|V_{\mathbf{q,k_i-q}}|^2 S(q) S(k_i-q)$ by a level-dependent
constant $n \lambda_n$, which represents the effective weight of the
memory kernel at level $n$. This brings the memory functions into the form
$K_n(t) \mapsto \lambda_n \phi_{n+1}(t)$. Finally, we assume that the density correlation functions
decay so slowly that the overdamped limit can be applied [$\ddot{\phi}_n(t)$ = 0].
Under these assumptions, we arrive at the generic schematic hierarchy of Eq.\ (\ref{eq:sGMCThierarchy}).
These schematic GMCT equations are microscopically motivated, but lack any explicit
$\mathbf{k}$-dependence. It is evident that Eq.\ (\ref{eq:sGMCThierarchy})
represents an infinite hierarchy of coupled equations, i.e., the time evolution
of any $\phi_n(t)$ is governed by $\phi_{n+1}(t)$, which in turn is governed by
$\phi_{n+2}(t)$, etc. Equation (\ref{eq:sGMCThierarchy}) is subject to the initial conditions
$\phi_n(0) = 1$ for all $n$, which follows directly from the definitions of
Eq.\ (\ref{eq:phinorm}).

The general solution of Eq.\ ({\ref{eq:sGMCThierarchy}) may be written in terms of
the Laplace transform as
\begin{equation}
\label{eq:phin_s}
\hat{\phi}_n(s) = \left( s + \frac{\mu_n}{1+\lambda_n \hat{\phi}_{n+1}(s)}
                  \right)^{-1}.
\end{equation}
In order to find a general expression for the relaxation time [Eq.\ (\ref{eq:taudef})],
we can iterate Eq.~(\ref{eq:phin_s})
$k$ times for $s=0$ to yield 
\begin{equation}
\label{eqa:phin_s=0it}
\tau_n \simeq \frac{1}{\mu_n} 
          \sum_{m=0}^{k} \prod_{i=0}^{m-1} \frac{\lambda_{n+i}}{\mu_{n+1+i}}. 
          + \left( \prod_{i=0}^{k} \frac{\lambda_{n+i}}{\mu_{n+i}} \right) 
          \tau_{n+k+1}.
\end{equation}
This equation may be further simplified by a suitable
choice of $\mu_n$ and $\lambda_n$ such that the second term will vanish for
$k \to \infty$, and we arrive at Eq.\ (\ref{eq:phin_s=0it}) of the main text.
Similarly, for the inverse plateau height we find by iteration
\begin{equation}
\label{eqa:qn_it}
\frac{1}{q_n} = \sum_{m=0}^k \prod_{i=0}^{m-1} \frac{\mu_{n+i}}{\lambda_{n+i}}
              + \left( \prod_{i=0}^{k} \frac{\mu_{n+i}}{\lambda_{n+i}} \right)
                \frac{1}{q_{n+k+1}},
\end{equation}
where again 
the second term vanishes in the limit of $k \rightarrow \infty$
if $\mu_n$ and $\lambda_n$ are chosen
appropriately [Eq.\ (\ref{eq:qn_it})].

\section{Infinite GMCT hierarchies with standard-MCT-like behavior}
\label{app:sharp}

Here we provide more mathematical details on the MCT-like infinite hierarchies
discussed in the main text.
We start with the hierarchy defined by
\begin{eqnarray}
\mu_n     &=& n \nonumber \\
\lambda_n &=& \Lambda(n+c),
\end{eqnarray}
with $c \geq 0$. As already mentioned in the main text, this type of
hierarchy exhibits an MCT-like transition at the critical point $\lambda_c=1$.
For $\Lambda>1$, the long-time limit of the
$2n$-point density correlator $\phi_n(t)$ satisfies
\begin{eqnarray}
\label{eq:qn_general_l(n+c)}
\frac{1}{q_n} &=& \sum_{m=0}^{\infty} 
                  \left( \prod_{i=1}^{m} \frac{n+i-1}{n+i-1+c} \right) 
                  \left( \frac{1}{\Lambda} \right)^m 
                  \nonumber \\
              &=& {}_2F_1(1,n;n+c; 1/\Lambda),
\end{eqnarray}
where ${}_pF_q(a_1, \hdots, a_p;b_1, \hdots, b_q; z) = \sum_{k=0}^\infty
\frac{(a_1)_k \dots (a_p)_k}{(b_1)_k \dots (b_q)_k} \frac{z^k}{k!}$ is the
generalized hypergeometric function with $(x)_k$ the Pochhammer symbol
\cite{abramowitz:64}. For $c=0$, this expression simplifies to
\begin{equation}
\frac{1}{q_n} = \sum_{m=0}^{\infty} \left( \frac{1}{\Lambda} \right)^m 
              = \left( 1-\frac{1}{\Lambda} \right)^{-1},
\end{equation}
which is independent of $n$. The case $c=1$ yields
\begin{equation}
\frac{1}{q_n} = \sum_{m=0}^{\infty} 
                 \left( \prod_{i=1}^{m} \frac{n+i-1}{n+i} \right) 
                 \left( \frac{1}{\Lambda} \right)^m 
              = \sum_{m=0}^{\infty} \frac{n}{n+m} 
                \left( \frac{1}{\Lambda} \right)^m,
\end{equation}
which for $n=1$ becomes
\begin{equation}
\frac{1}{q_1} = \sum_{m=0}^{\infty} \frac{1}{1+m} 
                \left( \frac{1}{\Lambda} \right)^m
              = \Lambda \log\left( \frac{\Lambda}{\Lambda-1} \right).
\end{equation}
Thus, for a hierarchy with $c=1$, the plateau height of the two-point density correlator
$\phi_1(t)$ grows logarithmically in the glassy regime.
Let us now consider the special case $c=\gamma$, where $\gamma \approx 1.76498$
is the standard-MCT exponent. As noted in the main text, this particular hierarchy exhibits a power-law
divergence of the relaxation time with exponent $\gamma$, similar to the predictions of the $F_2$ model.
The expression for the (inverse) plateau height can be found by employing the
series expansion of the hypergeometric function [Eq. (\ref{eq:qn_general_l(n+c)})], which yields
for $n=1$ and $\Lambda \gtrsim 1$
\begin{eqnarray}
\frac{1}{q_1} &=& \Gamma(\gamma+1) 
                \bigg\{
                (\Lambda-1)^{\gamma-1} \left[ \Gamma(1-\gamma) + \mathcal{O}(\Lambda-1) \right] +  \nonumber \\
              & & \frac{\Gamma(\gamma-1)}{[\Gamma(\gamma)]^2} + \mathcal{O}(\Lambda-1)
                \bigg\},
\end{eqnarray}
where $\Gamma(x)$ is the gamma function. Given that $1 < \gamma < 2$, we can
isolate the leading term by writing
\begin{eqnarray}
\frac{1}{q_1} 
              &=& \frac{\Gamma(\gamma+1) \Gamma(\gamma-1)}{[\Gamma(\gamma)]^2} \times \nonumber \\
              & &    \left[ 1 + \frac{\Gamma(1-\gamma) 
                  [\Gamma(\gamma)]^2}{\Gamma(\gamma-1)}(\Lambda-1)^{\gamma-1} 
                  + \mathcal{O}(\Lambda-1) \right]
              \nonumber \\
              &=& \frac{\gamma}{\gamma-1} 
                  \left[ 1 + 
                  \frac{\pi (\gamma-1)}{\sin(\pi \gamma)}(\Lambda-1)^{\gamma-1} 
                  + \mathcal{O}(\Lambda-1) \right]. \nonumber \\
              & & 
\end{eqnarray}
Inverting this equation yields
\begin{eqnarray}
q_1 &=& \frac{\gamma-1}{\gamma} \left[ 1 - 
        \frac{\pi (\gamma-1)}{\sin(\pi \gamma)}(\Lambda-1)^{\gamma-1} 
        + \mathcal{O}(\Lambda-1) \right]
    \nonumber \\
    &=& \frac{\gamma-1}{\gamma} \left[ 1 + 
        \frac{\pi (\gamma-1)}{\sin[\pi (\gamma-1)]}(\Lambda-1)^{\gamma-1} 
        + \mathcal{O}(\Lambda-1) \right]. \nonumber \\
   & &
\end{eqnarray}
Thus, for this type of hierarchy, the plateau height increases to leading order as
$(\Lambda-1)^{\gamma-1} \approx (\Lambda-1)^{0.765}$.

It is well established that the $F_2$ model predicts a long-time limit of $\phi_1(t)$
that behaves as
$q_1^{F_2} = \allowbreak{ (1+\sqrt{1-1/\Lambda})/2 }$ \cite{leutheusser:84}.
The Taylor series of $1/q_1^{F_2}$ is given by
\begin{equation}
\label{eq:TaylorqF2}
\frac{1}{q_1^{F_2}} = 1+ \sum_{m=1}^{\infty} \prod_{i=1}^m 
                      \left( \frac{2i-1}{2i+2} \right)
                      \left( \frac{1}{\Lambda} \right)^m,
\end{equation}
which can be compared with the general $1/q_1$ expression for an
arbitrary infinite GMCT hierarchy [Eq.\ (\ref{eq:qn_it})].
One may readily verify that an exact mapping between $q_1^{F_2}$ and
$q_1$ requires for the GMCT coefficients
\begin{equation} 
\label{eq:map_qF2}
\frac{\lambda_n}{\mu_n} = \frac{2n+2}{2n-1} \Lambda.
\end{equation}
A simple set of parameters satisfying this relation is $\mu_n=n$ and $\lambda_n
= \Lambda \frac{2n(n+1)}{2n-1}$. The $\alpha$-relaxation time associated with
this hierarchy is
\begin{eqnarray}
\tau_1 &=& \sum_{m=0}^{\infty} 
           \left( \prod_{i=1}^{m} \frac{2i}{2i-1} \right) \Lambda^m  
           \nonumber \\
       &=& \frac{ \sqrt{\Lambda} \arcsin{\sqrt{\Lambda}} + \sqrt{1-\Lambda}}{(1-\Lambda)^{3/2}},
\end{eqnarray}
which diverges for $\Lambda \rightarrow 1$ as $\tau_1 \sim (\pi/2) (1-\Lambda)^{-3/2}$.
Note this divergence is slightly different from that predicted
by standard MCT.

As a final example of $F_2$-like GMCT, we discuss the numerically fitted infinite hierarchy.
In this case, we seek to (numerically) reproduce both the relaxation-time and plateau-height
scaling of the $F_2$ model by a suitable choice of $\mu_n$ and $\lambda_n$.
Since Eq.\ (\ref{eq:map_qF2}) must hold, we can write for the relaxation time [see also
Eq.\ (\ref{eq:phin_s=0it})]
\begin{equation}
\label{eq:mun_exact_F2GMCT}
\tau_1 = \frac{1}{\mu_1} + \sum_{m=1}^{\infty} \frac{1}{\mu_{m+1}} 
                              \left( \prod_{i=1}^m \frac{2i+2}{2i-1} \right)
                              \Lambda^m.
\end{equation}
This series should be matched to the power law prediction of the $F_2$ scheme,
which diverges for $\Lambda \rightarrow 1$ as
\begin{eqnarray}
\label{eq:tauF2series}
\tau_1^{F_2} &=& \frac{A}{(1-\Lambda)^{\gamma}} 
               \nonumber \\
           &=& A\left[ 1 + \sum_{m=1}^{\infty} \left( \prod_{i=1}^m \frac{i+\gamma-1}{i} \right) \Lambda^m \right]
               \nonumber \\
           &\approx& 1 + A\sum_{m=1}^{\infty} \left( \prod_{i=1}^m \frac{i+\gamma-1}{i} \right) \Lambda^m,
\end{eqnarray}
The value of $A\approx 1.2573$ has been obtained from a numerical fit of the $F_2$ data
close to the transition ($\Lambda \lesssim 1$).
Instead of working with Eq.\ (\ref{eq:mun_exact_F2GMCT}) directly,
we now take the ansatz
\begin{equation}
\label{eq:ansatz}
\tau_1 = 1 + \sum_{m=1}^{\infty} \left( \prod_{i=1}^{m} \frac{i+a}{i+b} \right) \Lambda^m,
\end{equation}
and seek to match the $a$ and $b$ constants to Eq.\ (\ref{eq:tauF2series}).
In the limit of $m\rightarrow\infty$, we have
\begin{eqnarray}
\prod_{i=1}^{m-1} \frac{i+a}{i+b} &\sim& \frac{\Gamma(1+b)}{\Gamma(a+1)} m^{a-b}
                                         \nonumber \\
                                  &\sim& A \prod_{i=1}^{m-1} \frac{i+\gamma-1}{i},
\end{eqnarray}
which can be rewritten as
\begin{equation}
\frac{\Gamma(1+b)}{\Gamma(a+1)} m^{a-b} \sim A \frac{1}{\Gamma(\gamma)} m^{\gamma-1}.
\end{equation}
Setting $a=\gamma-1+b$ readily yields
\begin{equation}
A = \frac{\Gamma(\gamma)\Gamma(b+1)}{\Gamma(b+\gamma)},
\end{equation}
implying that $b\approx-0.23772$ and $a\approx 0.52726$. This set of constants thus enforces
that Eq.\ (\ref{eq:ansatz}) behaves as the MCT power law of Eq.\ (\ref{eq:tauF2series}).
A comparison between the ansatz (\ref{eq:ansatz}) and Eq.\ (\ref{eq:mun_exact_F2GMCT})
now gives for the $\mu_n$ coefficients
\begin{equation}
\frac{1}{\mu_{n+1}} \left( \prod_{i=1}^n \frac{2i+2}{2i-1} \right) = \prod_{i=1}^n \frac{i+a}{i+b}, 
\end{equation}
or, more explicitly,
\begin{equation}
\mu_n = \prod_{i=1}^{n-1} \frac{2i+2}{2i-1} \frac{i+b}{i+a}.
\end{equation}
The $\lambda_n$ parameters subsequently follow from Eq.\ (\ref{eq:map_qF2}), ensuring that
also the plateau height $q_1^{F_2}$ is correctly reproduced.

\section{Infinite GMCT hierarchies with avoided transitions}
\label{app:avoided}

In this Appendix, we focus on the relaxation-time behavior of infinite hierarchies
of the form $\mu_n=n$ and $\lambda_n=\Lambda n^{1-\nu}$, with $\Lambda,\nu > 0$. This class
of hierarchies lacks a sharp MCT-like transition for any finite $\Lambda$.
Starting with Eq.\ (\ref{eq:phin_s=0it}) of the main article, we have 
\begin{equation}
\tau_1 = \frac{1}{\mu_n} 
         \sum_{m=0}^{\infty} \prod_{i=0}^{m-1} \frac{\lambda_{n+i}}{\mu_{n+1+i}} 
       = \sum_{m=1}^{\infty} \frac{1}{\lambda_m} \prod_{i=1}^{m} \frac{\lambda_i}{\mu_i},
\end{equation}
which can be rewritten as
\begin{eqnarray}
\tau_1 &=& \sum_{m=1}^{\infty} \frac{1}{\Lambda m^{1-\nu}} \prod_{i=1}^{m} \frac{\Lambda i^{1-\nu}}{i}
           \nonumber \\
       &=& \frac{1}{\Lambda} \sum_{m=1}^{\infty} \frac{1}{[(m-1)!]^{\nu}} \frac{\Lambda^m}{m}
           \nonumber \\
       &=& \frac{1}{\Lambda} \sum_{m=1}^{\infty} \frac{1}{[(m-1)!]^{\nu}} \int_0^{\Lambda} x^{m-1} dx
           \nonumber \\
       &=& \frac{1}{\Lambda} \int_0^{\Lambda} dx \sum_{m=0}^{\infty} \frac{x^m}{(m!)^{\nu}}.
\end{eqnarray}
Introducing the function
\begin{equation}
f_{\nu}(x) = \sum_{m=0}^{\infty} \frac{x^m}{(m!)^{\nu}}
\end{equation}
now yields for the relaxation time
\begin{equation}
\label{eq:tau1_fvx}
\tau_1 = \frac{1}{\Lambda} \int_0^{\Lambda} f_{\nu}(x) dx.
\end{equation}
For the special case $\nu=1$ we thus find $f_1(x) = \sum_{m=0}^{\infty} x^m/(x!) =\exp(x)$ and
$\tau_1 = [\exp(\Lambda)-1]/\Lambda$, and for $\nu=2$ we have
 $f_2(x) = \sum_{m=0}^{\infty} x^m/(x!)^2 =I_0(2\sqrt{x})$ and $\tau_1 = I_1(2\sqrt{\Lambda})/\sqrt{\Lambda}$,
where $I_l$ is the modified Bessel function of the first kind of order $l$.

In order to find a closed expression for $\tau_1$ for arbitrary $\nu>0$, we
must evaluate the scaling behavior of $f_{\nu}(x)$ as a function of $\nu$. This
is a rather involved derivation; for simplicity we will focus on the asymptotic
behavior for $x \rightarrow \infty$ ($\Lambda\rightarrow\infty$).  We first
approximate the sum in $f_{\nu}(x)$ by an integral,
\begin{eqnarray}
f_{\nu}(x) &\sim& \int_{0}^{\infty} dm \frac{x^m}{\Gamma(m+1)^{\nu}}
                     \nonumber \\
           &=& \int_{0}^{\infty} dm \exp\left[ \log \left( \frac{x^m}{\Gamma(m+1)^{\nu}} \right) \right],
\end{eqnarray}
and expand the logarithm up to second order around the point $m=a$,
\begin{eqnarray}
\label{eq:expandlog}
\log & & \left[ \frac{x^m}{\Gamma(m+1)^{\nu}} \right] 
      = \log\left( x^a \Gamma(a+1)^{-\nu}\right) 
      + 
      \nonumber \\
      & & \left[ \log(x)-\nu \psi^{(0)}(a+1) \right] (m-a)
      \nonumber \\
      & & - (\nu/2) \psi^{(1)}(a+1) (m-a)^2 + \mathcal{O}(m-a)^3,
\end{eqnarray}
where $\psi^{(i)}(x)$ is the polygamma function of order $i$, i.e.\ the $(i+1)$-th derivative
of the logarithm of the gamma function. We will choose $a=x^{1/v}$
so that the linear term in Eq.\ (\ref{eq:expandlog}) vanishes.
Discarding the higher-order terms
and extending the range of $m-a$ to $\pm\infty$ yields
\begin{eqnarray}
f_{\nu}(x) &\sim& \frac{x^a}{\Gamma(a+1)^{\nu}} 
                     \int_{-\infty}^{\infty} dm \exp\left[ -(\nu/2)\psi^{(1)}(a+1) m^2 \right] 
                     \nonumber \\
           &=& \sqrt{\frac{2\pi}{\nu\psi^{(1)}(a+1)}} \frac{x^a}{\Gamma(a+1)^{\nu}},
\end{eqnarray}
where we have performed Gaussian integration.
For $a\rightarrow\infty$ we have, to leading order, $\Gamma(a+1) \approx a^a \exp(-a) \sqrt{2\pi a}$
and hence
\begin{equation}
f_{\nu}(x) \sim \frac{(2\pi a)^{(1-\nu)/2}}{\nu^{1/2}} \frac{\exp(\nu a) x^a}{a^{\nu a}}.
\end{equation}
Substituting $a=x^{1/\nu}$ finally yields
\begin{equation}
\label{eq:fvx_generalv}
f_{\nu}(x) \sim \frac{(2\pi)^{(1-\nu)/2}}{\nu^{1/2}} x^{(1-\nu)/2\nu} \exp(\nu x^{1/\nu}).
\end{equation}
One may verify that, for the special cases $\nu=1$ and $\nu=2$, Eq.\ (\ref{eq:fvx_generalv})
indeed describes the asymptotic behavior of $f_1(x)$ and $f_2(x)$.

Finally, we seek to obtain a closed expression for the relaxation time $\tau_1$ for general $\nu>0$.
Substituting Eq.\ (\ref{eq:fvx_generalv}) into Eq.\ (\ref{eq:tau1_fvx}) and
setting $y=\nu x^{1/\nu}$ gives
\begin{equation}
\tau_1 \sim (2\pi)^{(1-\nu)/2} \nu^{-\nu/2} \frac{1}{\Lambda} 
         \int_{0}^{\nu \Lambda^{1/\nu}} dy y^{(\nu-1)/2} \exp(y),
\end{equation}
which, by changing variables $y \rightarrow \nu\Lambda^{1/\nu}-y$, can be rewritten as
\begin{eqnarray}
\tau_1 &\sim& \frac{(2\pi)^{(1-\nu)/2}}{\nu^{1/2}} \Lambda^{-(\nu+1)/2\nu} \exp(\nu\Lambda^{1/\nu}) 
       \times \nonumber \\
         & & \int_{0}^{\nu \Lambda^{1/\nu}} dy \left(1-\frac{y}{\nu \Lambda^{1/\nu}} \right)^{(\nu-1)/2} \exp(-y).
\end{eqnarray}
Expanding the power-law factor in the integrand and extending the integration range to $+\infty$
finally yields, to leading order,
\begin{equation}
\tau_1 \sim \frac{(2\pi)^{(1-\nu)/2}}{\nu^{1/2}} \Lambda^{-(\nu+1)/2\nu} \exp({\nu \Lambda^{1/\nu}}),
\end{equation}
which is equivalent to Eq.\ (9) of the main article. Again it may be verified that this equation,
which holds asymptotically for $\Lambda \gg 1$, is consistent with the expressions for
$\nu=1$ and $\nu=2$ given earlier.


\end{document}